\documentclass[letterpaper, 10 pt, conference]{ieeeconf}  

\IEEEoverridecommandlockouts                              

\usepackage{amsfonts, dsfont, mathtools, amsmath, amssymb, bbold, mathrsfs, bm, upgreek}
\usepackage{graphicx, float, epsfig, color, psfrag, adjustbox, enumerate}
\usepackage{subcaption}
\usepackage{refcount, url}
\usepackage[noadjust]{cite}
\usepackage[usenames,dvipsnames,svgnames,table]{xcolor}
\usepackage{algorithm}
\usepackage{algpseudocode}

\algrenewcommand\alglinenumber[1]{#1:}
\usepackage{multirow}
\usepackage{multicol}
\usepackage{tikz}
\usetikzlibrary{positioning,arrows}

\title{\LARGE \bf
Energy Shaping Control of a CyberOctopus Soft Arm
}

\author{Heng-Sheng Chang$^{1,2}$, Udit Halder$^{2}$, Chia-Hsien Shih$^{1}$, Arman Tekinalp$^{1}$, Tejaswin Parthasarathy$^{1}$,\\ [5pt] Ekaterina Gribkova$^{3}$, Girish Chowdhary$^{2,4}$, Rhanor Gillette$^{3, 5}$, Mattia Gazzola$^{1,6,7}$, Prashant G. Mehta$^{1,2}$
\thanks{We gratefully acknowledge financial support from ONR MURI N00014-19-1-2373 (G.C., M.G, M.G.P.), NSF/USDA $\#$2019-67021-28989 (G.C., M.G.), and NSF EFRI C3 SoRo $\#$1830881 (M.G.). We also acknowledge computing resources provided  by the Blue Waters project (OCI- 0725070, ACI-1238993), a joint effort of the University of Illinois at Urbana-Champaign and its National Center for Supercomputing Applications, and the Extreme Science and Engineering Discovery Environment (XSEDE) Stampede2 (ACI-1548562) at the Texas Advanced Computing Center (TACC) through allocation TG-MCB190004.}%
\thanks{$^{1}$Department of Mechanical Science and Engineering, $^{2}$Coordinated Science Laboratory,
$^{3}$Neuroscience Program, $^{4}$Department of Agricultural and Biological Engineering, $^{5}$Department of Molecular and Integrative Physiology, $^{6}$National Center for Supercomputing Applications, \&  $^{7}$Carl R. Woese Institute for Genomic Biology, University of Illinois at Urbana-Champaign. Corresponding e-mail:
        {\tt\small udit@illinois.edu}}%
}

\def\R{{\mathds{R}}}

\def\0{{\mathbb{0}}}
\def\1{{\mathds{1}}}

\def\a{{\mathbf{a}}}
\def\b{{\mathbf{b}}}

\newcommand{\inner}[2]{\left\langle #1, #2 \right\rangle}

\definecolor{db}{RGB}{23,20,119}
\definecolor{dg}{RGB}{2,101,15}

\newtheorem{proposition}{Proposition}[section]

\newtheorem{remark}{Remark}

\usepackage{upgreek}
\newcommand{\dif}{\mathrm{d}}
\newcommand{\transpose}{\intercal}
\newcommand{\set}[1]{\left\{#1\right\}}
\newcommand{\defined}{\coloneqq}

\newcommand{\material}[1]{
	\ifthenelse{\equal{#1}{\kappa}}{\upkappa}{
	\ifthenelse{\equal{#1}{\nu}}{\upnu}{
	\ifthenelse{\equal{#1}{\omega}}{\upomega}{
	\ifthenelse{\equal{#1}{\sigma}}{\upsigma}{
	\ifthenelse{\equal{#1}{\theta}}{\uptheta}{
	\mathsf{#1}}}}}}
}

\newcommand{\target}{*}
\newcommand{\intrinsic}{\circ}

\newcommand{\deformations}{w}
\newcommand{\states}{q}
\newcommand{\momentums}{p}
\newcommand{\costates}{\lambda}
\newcommand{\Hamiltonian}{\mathcal{H}}
\newcommand{\potential}{\mathcal{V}}
\newcommand{\kinetic}{\mathcal{T}}
\newcommand{\mass}{M}
\newcommand{\controlHamiltonian}{H}

\renewcommand{\a}{\mathsf{a}}
\renewcommand{\b}{\mathsf{b}}

\tikzstyle{block} = [draw,  rectangle, minimum height=3em, minimum width=6em]
\tikzstyle{sum} = [draw, circle, node distance=1cm]
\tikzstyle{input} = [coordinate]
\tikzstyle{output} = [coordinate]
\tikzstyle{pinstyle} = [pin edge={to-,thin,black}]
\graphicspath{{figures/}}

\begin{document}
\bstctlcite{BSTcontrol} 
\maketitle
\thispagestyle{empty}
\pagestyle{empty}


\begin{abstract}
This paper entails the application of the energy shaping methodology to control a flexible, elastic Cosserat rod model.  Recent interest in such continuum models stems from applications in soft robotics, and from the growing recognition of the role of mechanics and embodiment in biological control strategies: octopuses are often regarded as iconic examples of this interplay. The dynamics of the Cosserat rod, here modeling a single octopus arm, are treated as a Hamiltonian system and the internal muscle actuators are modeled as distributed forces and couples. The proposed energy shaping control design procedure involves two steps: (1) a potential energy is designed such that its minimizer is the desired equilibrium configuration; (2) an energy shaping control law is implemented to reach the desired equilibrium.  By interpreting the controlled Hamiltonian as a Lyapunov function, asymptotic stability of the equilibrium configuration is deduced.  The energy shaping control law is shown to require only the deformations of the equilibrium configuration.  A forward-backward algorithm is proposed to compute these deformations in an online iterative manner.  The overall control design methodology is implemented and demonstrated in a dynamic simulation environment.  Results of several bio-inspired numerical experiments involving the control of octopus arms are reported.

\end{abstract}

\begin{keywords}
Cosserat rod, Hamiltonian systems, energy-shaping control, soft robotics, octopus
\end{keywords}

\section{Introduction} \label{sec:intro}
In recent years, the octopus has become an iconic example of the potential opportunities that lie in the use of soft, compliant materials in robotic applications, to enhance dexterity, safety, and body reconfiguration abilities \cite{Laschi:2012, min2019softcon}. Indeed, the octopus and other soft-bodied animals are able to coordinate virtually infinite degrees of freedom into a rich repertoire of complex manipulation and motion patterns, from reaching, grasping, fetching, to crawling and swimming \cite{sumbre2001control, sumbre2005motor, levy2017motor}. Recent proof-of-concept soft robots continue to highlight the need for theoretical and algorithmic control approaches that are specifically tailored to such distributed and compliant mechanical systems. 
This provides the motivation for the work reported in this paper where we apply energy shaping control techniques to control a virtual octopus arm.

The dynamics of the arm are modeled using the Cosserat theory of elastic rods~\cite{antman1995nonlinear}.
In contrast to typical rigid link models of classical robotics, Cosserat rod models capture, through linear and angular momentum balances, the (one-dimensional) continuum and distributed nature of elastic slender bodies deforming in space. These models account for all modes of deformation -- bend, twist, stretch, shear -- induced by external and internal forces and couples. 

Our control-oriented viewpoint is to interpret the rod as a Hamiltonian system \cite{simo1988hamiltonian, dichmann1996hamiltonian} where the potential energy is expressed in terms of strains.  This allows us to apply an energy shaping control design procedure that involves two steps: (1) a potential energy is designed such that its minimizer is the desired static equilibrium (encoding the octopus' goal, e.g., reaching an object);  (2) an energy shaping control law is implemented to achieve the desired equilibrium.  In a standard manner, by interpreting the controlled Hamiltonian as a Lyapunov function, the equilibrium configuration is shown to be asymptotically stable. 
The energy shaping control methodology has a rich history in robotics~\cite{van2000l2, ortega2001putting}.  Apart from our work, this method has recently been applied to the control of soft manipulators based upon a finite dimensional rigid link model~\cite{franco2020energy}. 
The proposed procedure has several useful features. It yields a simple closed-form formula for the control law which is easily integrated in a realistic simulation.  The modified potential energy and the controlled Hamiltonian have useful physical interpretations as modified stress-strain relationships. Our simple control law provides a benchmark for more sophisticated forms of controls where additional constraints due to sensing and actuation may be taken into account. 
The algorithms described in this paper are demonstrated in a computational {\em CyberOctopus} which is being developed to simulate soft body mechanics coupled with distributed sensory-motor infrastructure operating in a realistic physical environment.  The mechanics component of the \textit{CyberOctopus} is simulated with \textit{Elastica}, an existing software for the numerical modeling and simulation of Cosserat rods~\cite{gazzola2018forward, zhang2019modeling} in 3D space.  
Several reaching motion patterns inspired by results reported in the octopus' literature are demonstrated in numerical experiments.

\begin{figure*}[!ht]
\vspace{5pt}
\centering
	\includegraphics[width=\textwidth]{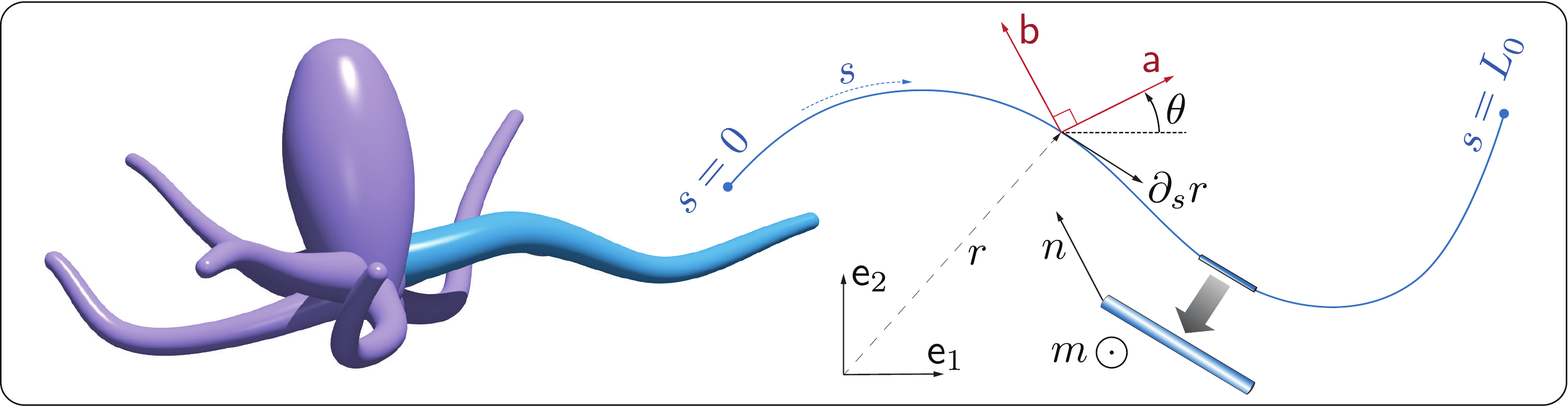}
	\caption{Modeling an octopus arm as a special Cosserat rod}
\label{fig:rod}
\vspace{-15pt}
\end{figure*}

The outline of the remainder of this paper is as follows. The static and dynamic equations of the classical planar Cosserat rod theory are introduced in Sec.~\ref{sec:model}.  The section includes a self-contained discussion of an optimal control-type formulation of the rod statics, and the Hamiltonian formulation of the rod dynamics.  The proposed energy-shaping control design procedure appears in Sec.~\ref{sec:control}.  The details of the simulation platform and the results of the numerical experiments appear in Sec.~\ref{sec:numerics} and Sec.~\ref{sec:experiments}, respectively. The conclusions and directions for future research are briefly described in Sec.~\ref{sec:conclusion}.



\section{Cosserat rod model of a single arm} \label{sec:model}

Let $\set{\mathsf{e}_1,\mathsf{e}_2}$ denote a fixed orthonormal basis for the two-dimensional lab frame\footnote{Although all the considerations of this paper are applicable to the general three-dimensional (3D) Cosserat rod models, we provide the exposition for two-dimensional (2D) models.  The notation is simpler and the key ideas are  communicated more easily to a broader audience.}. In its reference undeformed configuration, the rod is of length $L_0$ and lies parallel to the $\mathsf{e}_1$ axis. The independent coordinates are time $t \in \R$ and the arc-length of the centerline $s \in [0, L_0]$. The partial derivatives with respect to $t$ and $s$ are denoted as $\partial_t$ and $\partial_s$, respectively. The \textit{state} of the rod is described by the vector-valued function $\states$  (Fig.~\ref{fig:rod})
	\begin{equation}
		\states(s,t)\defined\begin{bmatrix}x(s,t)\\y(s,t)\\ \theta(s,t)\end{bmatrix} \nonumber
	\end{equation}
	where $r=(x,y)\in\R^2$ denotes the position vector of the centerline, and the angle $\theta\in\R$ defines a material frame spanned by the orthonormal pairs $\{\a,\b\}$, where $\a = \cos \theta \,\mathsf{e}_1 + \sin \theta \, \mathsf{e}_2, ~ \b = -\sin \theta \, \mathsf{e}_1 + \cos \theta \, \mathsf{e}_2$.
The vector $\a$ is normal to the cross section:  it captures the shear deformations whereby the cross section `shears' relative to the tangent of the centerline.  


\subsection{Statics -- an optimal control viewpoint} \label{sec:statics}

The statics of the rod require consideration of the rod's potential energy denoted as $\potential$.  It is a functional of the strains, i.e., curvature, stretch and shear. Strains are related to the local frame $\{\a, \b\}$ through $\partial_s r =\nu_1\a+\nu_2\b$, where $\nu_1$ and $\nu_2$ represent stretch and shear, respectively. The curvature $\kappa := \partial_s
\theta$ completes the triad of deformations $\deformations:=(\nu_1,\nu_2,\kappa)$ that fully characterizes the rod's kinematics
\begin{equation}
		\partial_s\states= {f}(\states,\deformations):=
			\begin{bmatrix}
				\nu_1\cos\theta-\nu_2\sin\theta\\
				\nu_1\sin\theta+\nu_2\cos\theta\\
				\kappa
			\end{bmatrix}
			\label{eq:state_kinematics}
\end{equation}
and potential energy
\begin{align}
\potential(\states) &= \int_0^{L_0} W\left(\deformations (s)\right) \dif s \nonumber
\end{align}
where $W : (\nu_1, \nu_2, \kappa) \mapsto \R$ is the
energy stored in the rod because of its mechanical deformation. Under the assumption of a perfectly elastic material characterized by a linear stress-strain relation, $W$ takes the quadratic form
	\begin{equation}
		W = \frac{1}{2} \left( EA(\nu_1-\nu_1^\intrinsic)^2 + GA(\nu_2-\nu_2^\intrinsic)^2+ EI(\kappa-\kappa^\intrinsic)^2 \right)
		\label{eq:stored_energy}
	\end{equation}
	where $E$ and $G$ are the material Young's and shear moduli, $A$ and $I$ are the cross sectional area and second moment of area, and $\nu_1^\intrinsic$, $\nu_2^\intrinsic$, $\kappa^\intrinsic$ are the intrinsic deformations that determine the rod's shape at rest. If $\nu_1^\intrinsic \equiv 1$, $\nu_2^\intrinsic \equiv 0$, $\kappa^\intrinsic \equiv 0$, then the rest configuration is a straight rod of length $L_0$.

The statics of the rod admit an interesting optimal control
re-formulation; c.f.,~\cite{bretl2014quasi, till2017elastic}. Any static configuration of
the rod is a stationary point of the potential energy $\potential$ with the constraint expressed by~\eqref{eq:state_kinematics}
\begin{align} \label{eq:optimal_control_fixed_end}
	\underset{\deformations(\cdot)}{\text{minimize}} \quad  & \potential = \int_0^{L_0} W(\deformations (s))~ \dif s, \\
	\text{ subject to }~ \partial_s\states=&{f}(\states,\deformations), ~\text{ with }\states(0)=\states_0, ~ \states(L_0) = \states_1 \nonumber
\end{align}
Here, $\states_0$ and $\states_1$ are the states of the rod at the base ($s=0$) and at the tip ($s=L_0$). Desired static configurations of the rod are obtained via the application of the Pontryagin's Maximum
Principle (PMP).
Write the control Hamiltonian as
	\begin{align}
		\controlHamiltonian (\states, \costates, \deformations) = \costates^{\transpose} {f} (\states, \deformations) - W(\deformations)
		\label{eq:control_hamiltonian}
	\end{align}
	where $\costates(s) = (\lambda_1(s), \lambda_2(s), \lambda_3(s))^{\transpose} \in \R^3$ is the costate vector.
The Hamilton's equations are given by
\begin{align}
\partial_s\states=&\frac{\partial \controlHamiltonian}{\partial \costates}  = {f}(\states,\deformations) \label{eq:states_evolution}\\
\partial_s \costates = &- \frac{\partial \controlHamiltonian}{\partial
                       \states}
				               = \begin{bmatrix}
0 \\ 0 \\ \big\lbrace -\nu_1 (-\lambda_1 \sin \theta + \lambda_2 \cos \theta) \\ ~~+  \nu_2 (\lambda_1 \cos \theta + \lambda_2 \sin \theta) \big\rbrace
\end{bmatrix}
\end{align}
The optimal deformations are obtained by pointwise maximization of the Hamiltonian~\eqref{eq:control_hamiltonian}.  For the quadratic choice of the stored energy function $W$, the maximization yields
	\begin{align}
	\begin{bmatrix}	EA(\nu_1 - 1) \\ GA \nu_2 \\ EI \kappa	\end{bmatrix} = \begin{bmatrix} \lambda_1 \cos \theta + \lambda_2 \sin \theta \\
	- \lambda_1 \sin \theta + \lambda_2 \cos \theta \\ \lambda_3
	\end{bmatrix}
	\label{eq:optimal_deformations}
	\end{align}

\begin{remark}
In the Cosserat rod theory, Eq.~\eqref{eq:states_evolution}-\eqref{eq:costates_evolution} are the well known static equations.  The costate variables $(\lambda_1,\lambda_2)$ and $\lambda_3$ represent, respectively, the internal forces and couple in the laboratory frame.  In the material frame, the internal forces and couple are denoted as $(n_1, n_2, m) := (\lambda_1 \cos \theta +
\lambda_2 \sin \theta, -\lambda_1 \sin \theta + \lambda_2 \cos \theta,
\lambda_3)$. Equation~\eqref{eq:optimal_deformations} provides a
relationship between the deformations and these internal forces and couple. More generally, 
\begin{align*}
n_i = \frac{\partial W}{\partial \nu_i}, i=1,2, ~~~m = \frac{\partial W}{\partial \kappa} 
\end{align*}
are referred to as the {\em  constitutive laws} or the load-strain relationships that characterize the material of the rod.

\end{remark}



\subsection{Dynamics -- the Hamiltonian form}

In a dynamic setting, the state $\states = (x, y, \theta)$ is
a function of both $s$ and $t$.  Let $\momentums = \mass \partial_t
\states$ denote the momentum, where $\mass = \text{diag}(\rho A,
\rho A, \rho I)$ is the inertia matrix and $\rho$ is the material density.
The kinetic energy is expressed as
\begin{align*}
\kinetic= \frac{1}{2} \int_0^{L_0} \left(\rho A ( (\partial_t x)^2 + (\partial_t y)^2) +  \rho I (\partial_t \theta)^2 \right) \dif s
\end{align*}
The Hamiltonian is the total energy of the system, $\Hamiltonian(\states, \momentums) = \kinetic(\momentums) + \potential(\states)$.

In the absence of external forces and couples, the dynamics of the rod are described by Hamilton's equations of classical mechanics
\begin{align}
\begin{split}
\frac{\dif  \states}{\dif t} &= \frac{\delta \Hamiltonian}{\delta \momentums} = \mass^{-1} \momentums \\
\frac{\dif  \momentums}{\dif t} &= - \frac{\delta \Hamiltonian}{\delta \states} = -\frac{\delta \potential}{\delta \states}
\end{split}
\label{eq:hamilton_equations}
\end{align}

The evolution equation~\eqref{eq:hamilton_equations} requires the
specification of boundary conditions at $s=0$ and $s=L_0$ as well as
initial conditions at $t=0$.  These together with the explicit form of
the dynamic equations of the rod, appear in Sec.~\ref{sec:numerics}. 

\section{Control design} \label{sec:control}
The Hamiltonian control system is expressed as
\begin{align*}
	\begin{split}
		\frac{\dif \states}{\dif t} &= \frac{\delta \Hamiltonian}{\delta \momentums}  \\
		\frac{\dif \momentums}{\dif t} &= - \frac{\delta \Hamiltonian}{\delta \states} + \mathsf{G} (q,p) u
	\end{split}
\end{align*}
In an octopus, the control term $\mathsf{G} (q,p) u$ represents the distributed forces and torques generated by various kinds of muscles, e.g. longitudinal, transverse, and oblique muscles.  
In this paper, we take $\mathsf{G}(q,p)$ to be the identity, inferring that forces and torques of any magnitude and direction can be produced by the control vector $u$. Modeling of realistic anatomy, geometry, and mechanics of the internal muscle architecture is the subject of ongoing work.  

The control objective is to design a feedback control law for $u$ to manipulate the arm to perform a variety of control tasks: (i) displace and stabilize the tip of the rod ($s=L_0$) to a specified target location $\states^\target\in\R^3$ in an environment with obstacles; (ii) wrapping the arm around an object in order to grab it.  These objectives are closely inspired by the specific control behaviors observed in octopus arm movements. 
\color{black}

\subsection{Energy shaping control law}

The idea is to shape the potential energy of the rod, using techniques from the port-Hamiltonian control theory~\cite{takegaki1981new, ortega2001putting, van2000l2}. For this purpose, suppose one can design a potential energy, denoted as $\potential^d$, whose minimizer (static equilibrium) achieves the desired control objective\footnote{The design of the desired potential energy is the subject of the following sub-section.}.  Then the following proposition gives an explicit form of the control law:

\medskip

\begin{proposition} \label{prop:stabilization}
Let $\potential^d(\states)$ denote a desired potential energy function with minimum at a configuration $\bar{\states}$.  Then the control law
\begin{align}
u =- \frac{\delta}{\delta \states} (\potential^d - \potential) - \gamma \mass^{-1}\momentums, ~~~ \gamma >0
\label{eq:energy_shaping_control}
\end{align}
renders the point $(\bar{\states}, 0)$ asymptotically stable.
\end{proposition}

\medskip

A sketch of the proof (adapted from~\cite{takegaki1981new}) is provided next.  
The control law \eqref{eq:energy_shaping_control} serves to modify the
potential energy of the system to $\potential^d$
\begin{equation}
\frac{\dif \states}{\dif t} = \frac{\delta \bar{\Hamiltonian}}{\delta \momentums},\quad\frac{\dif \bar{\momentums}}{\dif t} = - \frac{\delta \bar{\Hamiltonian}}{\delta \states} - \gamma \mass^{-1} \momentums
\label{eq:hamilton_equations_controlled}
\end{equation}
where
\begin{align*}
\bar{\Hamiltonian} (\states, \momentums) = \kinetic(\momentums) + \potential^d(\states)
\end{align*}
is the modified control Hamiltonian.  Now, $\bar{\Hamiltonian} (\states, \momentums) \geq 0$ for all $(\states, \momentums)$, $\bar{\Hamiltonian} = 0$ only at $(\bar{\states}, 0)$ and along a solution trajectory of \eqref{eq:hamilton_equations_controlled} we have
\begin{align*}
\frac{\dif \bar{\Hamiltonian}}{\dif t} = - \gamma \inner{\frac{\dif \states}{\dif t}}{\frac{\dif \states}{\dif t}} \leq 0
\end{align*}
where the inner product above is taken in the $L^2$ sense. This shows that the total energy of the system is non-increasing.  By an application of the LaSalle's theorem, the solution converges to the largest invariant subset of $\set{(\states, \momentums) ~|~ \frac{\dif \bar{\Hamiltonian}}{ \dif t} = 0 }$ which is $(\bar{\states}, 0)$.  A rigorous application of LaSalle principle also requires one to show that the trajectories of the nonlinear semigroup of \eqref{eq:hamilton_equations_controlled} are precompact or relatively compact in an appropriate function space. This remains to be verified.       


\medskip

\begin{remark}
A justification of the small dissipation term in~\eqref{eq:hamilton_equations_controlled} can be provided in variety of ways, for example it can be physically assimilated to material viscoelastic effects.
\end{remark}

\medskip

It remains to determine the desired potential energy.  This is the subject of the next section.

\subsection{Design of desired potential energy} \label{sec:desired_potential_energy}
In order to design the desired potential energy, we build upon the optimal control re-formulation of the rod statics~\eqref{eq:optimal_control_fixed_end}. 
Specifically, we consider the following modified version of the problem:
\begin{align}
		\hspace{-5pt}
		\begin{split}
			\underset{\deformations(\cdot)}{\text{minimize}}   ~~& {\sf J} =
			\int_0^{L_0}  W(\deformations (s))+ \mu_{\text{grasp}}(s)\Phi_{\text{grasp}}(\states(s)) ~\dif s  \\
			   & \quad~~+\mu_{\text{tip}} \Phi_{\text{tip}}(\states(L_0), \states^\target) \\[5pt]
			\text{ subject to } ~& \partial_s\states={f}(\states,\deformations), ~ \text{ with }\states(0)=\states_0, ~ \states (L_0) ~ \text{free};\\
			\text{and}~~ & \Psi_j(\states)\leq 0, \quad j = 1, 2, \dots, N_{\text{obs}}
		\end{split}
		\label{eq:optimal_control_free_end}
\end{align}
With $\mu_{\text{grasp}} = \mu_{\text{tip}}=0$, $N_{\text{obs}}=0$ and a prescribed $q(L_0)$, this problem reduces to the original problem~\eqref{eq:optimal_control_fixed_end}.  In the control settings of this paper, these are chosen to satisfy various types of control objectives:
\begin{enumerate}
\item If there are obstacles in the environment, these are described by the state inequality constraint $\Psi_j(q) \leq 0$.
\item The terminal cost function $\Phi_{\text{tip}}(\cdot)$ is used to penalize the deviation of the arm tip from a specified target point $\states^\target$; $\mu_{\text{tip}}$ is a non-negative regularization parameter.  Such a model is useful for example to mimic an octopus arm reaching a prey in its environment. 
\item The state-dependent running cost function $\Phi_{\text{grasp}}(\cdot)$ and the weight function $\mu_{\text{grasp}}(\cdot)$ are motivated by the grasping control task.  In performing this task, a portion of the octopus arm wraps around and grasps an object in the environment.  
\end{enumerate} 
The regularization parameter $\mu_{\text{tip}}$ and the weight function $\mu_{\text{grasp}}$ are designed according to the underlying task; a representative guide is provided in Table \ref{tab:design}.  Additional details including explicit formulae for the functions $\Phi_{\text{tip}}, \Phi_{\text{grasp}}, \Psi$, and $\mu_{\text{grasp}}$ used in this work appear in Sec. \ref{sec:experiments}.

Following~\cite{mekarapiruk1997optimal}, the constrained optimal control problem \eqref{eq:optimal_control_free_end} is solved by augmenting the states $\states$ with $N_\text{obs}$ additional states, denoted as $\hat{q}_j$ for $j=1,\hdots,N_{\text{obs}}$.  The model of each additional state is defined as
	\begin{equation}
		\partial_s \hat{\states}_j=c_j(\states)=\max(\Psi_j(\states),0),\quad \hat{\states}_j(0)=0 \nonumber
	\end{equation}
	Note that $c_j(\states)$ is non-negative for each $j$. The terminal value $\hat{q}_j(L_0)$ is referred to as the performance index.  It indicates the degree to which the $j$-th inequality constraint has been violated along the length of the rod.  To minimize the performance index, the terminal cost function is modified as
	\begin{equation}
		\hat\Phi(\states(L_0),\hat{\states}(L_0))= \mu_{\text{tip}}\Phi_{\text{tip}}(\states(L_0), \states^\target) + \sum_{j=1}^{N_\text{obs}}\xi_j\hat{\states}_j(L_0) \nonumber
	\end{equation}
	where $\xi_j>0$ is the weight for the performance index $\hat{\states}_j(L_0)$.

	The Hamilton's equations of the state and the pointwise maximization condition for the Hamiltonian are exactly the same as before.  The equations for costates are modified to now also include additional terms on account of the constraints
	\begin{equation}
		\partial_s\costates=-\frac{\partial \hat{\controlHamiltonian}}{\partial\states}+\sum_{j=1}^{N_\text{obs}}\xi_j\frac{\partial c_j(\states)}{\partial\states}  = : g(s, q, \lambda, \deformations)
		\label{eq:costates_evolution}
	\end{equation}
where the modified control Hamiltonian $\hat{\controlHamiltonian}$ is written as 
\begin{align}
\hat{\controlHamiltonian}(s, \states, \costates, \deformations) = H(\states, \costates, \deformations) - \mu_{\text{grasp}}(s) \Phi_{\text{grasp}}(\states)
\end{align}
 The costate equation \eqref{eq:costates_evolution} is accompanied by the transversality condition 
\begin{align}\label{eq:transversality_condition}
\hspace*{-7pt}
\costates (L_0) = -\frac{\partial \hat{\Phi}}{\partial \states} (\states(L_0), \hat{\states}(L_0)) =  - \mu_{\text{tip}} \frac{\partial \Phi_{\text{tip}} }{\partial \states} (\states(L_0), \states^*)
\end{align}

Suppose the problem~\eqref{eq:optimal_control_free_end} is solved to obtain the static solution $\bar{\states}$. Then one possible approach to determine the desired potential energy function $\potential^d$ is as follows: 
	\begin{align}
		\potential^d(\states) &= \frac{1}{2} \int_0^{L_0} \left( EA (\nu_1 - \bar{\nu}_1 )^2 + GA (\nu_2 - \bar{\nu}_2)^2  \right. \nonumber \\
					& \left. \qquad \qquad \qquad + EI (\kappa - \bar{\kappa})^2 \right) \dif s
		\label{eq:desired_potential}
	\end{align}
	where $(\bar{\nu}_1,\bar{\nu}_2,\bar{\kappa})$ represent the optimal deformations corresponding to the solution $\bar{\states}$ of the control problem \eqref{eq:optimal_control_free_end}.  The quadratic formula may be replaced by any positive definite functional such that $\potential^d(\states) > 0$ for all $\states \neq \bar{\states}$, and $\potential^d(\bar{\states}) = 0$. 

	Using this choice yields the following explicit form of the energy shaping control law
	\begin{align}
		u &= -\begin{bmatrix}
					\frac{\partial}{\partial s }\left(
						\begin{pmatrix} \cos \theta  & - \sin \theta \\ \sin \theta & \cos \theta\end{pmatrix}
						\begin{pmatrix} EA (\bar{\nu}_1 -1) \\ GA \bar{\nu}_2 \end{pmatrix}  \right) \\
					{\partial_s (EI  \bar{\kappa})} + GA{\nu}_1 \bar{\nu}_2 - EA {\nu}_2 (\bar{\nu}_1 -1)
	\end{bmatrix} - \gamma \partial_t \states
	\label{eq:control}
	\end{align}

	Physically, this procedure is akin to artificially replacing the intrinsic strains of (\ref{eq:stored_energy}) with the optimal deformations $(\bar{\nu}_1,\bar{\nu}_2,\bar{\kappa})$. The energy shaping form of the controlled Hamiltonian dynamics generates the control inputs (which may be interpreted as muscle forces and couples) to bring the rod to its new equilibrium configuration.

%

\color{black}

\color{black}

\subsection{Algorithm}

	In summary, the proposed design procedure involves two steps: (i) In Step 1, the static deformations are obtained by solving the optimization problem~\eqref{eq:optimal_control_free_end}; (ii) In Step 2, the energy shaping dynamic control law \eqref{eq:control} is implemented to achieve the desired deformation.

	There are a number of ways to numerically solve the Hamilton's equations. Offline approaches include the use of a shooting method to solve the two point boundary value problem (BVP) \cite{morrison1962multiple}, or using continuation techniques \cite{healey2006straightforward}.  Once the optimal deformations are obtained, the control law is implemented directly using~\eqref{eq:control}.

Envisioning the control of a \textit{CyberOctopus} which interacts with a dynamic environment, an online approach is more appropriate. In this case, Step~1 is implemented to solve the BVP iteratively, interspacing every iteration with Step~2 directly within the simulation of the dynamic model. For the iterative solution of the BVP problem, a gradient ascent algorithm is used to update (or learn) the optimal deformations as follows
\begin{align} \label{eq:update_deformations}
		\frac{\dif\deformations}{\dif t}=\eta(t) \frac{\partial \hat{\controlHamiltonian}}{\partial \deformations} \left(s, \states, \costates, \deformations \right)
	\end{align}
where $\eta(t)$ is the update stepsize (or learning rate). For each $t$, the states are integrated forward from the initial condition $\states_0$, while the costates are integrated backward from the terminal condition that depends on the objective (see Table \ref{tab:design}. This algorithm is known in literature as forward-backward algorithm for optimal control~\cite{mcasey2012convergence, taghvaei2017regularization}, and is presented in Algorithm \ref{alg:forward_backward}. Convergence results typically require sufficiently small values of the step size update $\eta_k \Delta t$.    

	\begin{table}[tb]
		\centering
		\vspace{6pt}
		\caption{Design of Parameters in \eqref{eq:optimal_control_free_end}}
		\hskip-6pt
		\normalsize
		\renewcommand{\arraystretch}{1.2}
		\begin{tabular}{c||c|c}
		\hline
		Task & $\mu_{\text{grasp}}(s) $ & $\mu_{\text{tip}}$ \\
			\hline \hline
		Reaching with the tip, &	
		\multirow{2}{*}{0} &  \multirow{2}{*}{$\mu_{\text{tip}}>0$} \\
		w/ or w/o obstacles & &  \\ \hline
		\multirow{3}{*}{Grasping an object} &  \multirow{3}{*}{\shortstack[c]{Any non-negative\\ piecewise continuous \\ function of $s$}} & \multirow{3}{*}{$\mu_{\text{tip}} = 0$} 	\\ 
		 & &  \\
		 & & \\
		  	\hline

		\end{tabular}
		\label{tab:design}
	     \color{black}
	\end{table}

	\begin{algorithm}[t]
		\begin{algorithmic}[1]
			\Require Task (reaching, grasping etc.)
			\Ensure Optimal deformations $\bar{\deformations}=(\bar{\nu}_1,\bar{\nu}_2,\bar{\kappa})$
			\State Initialize: deformations $\deformations^{(0)}$, states at base ($s=0$) $\states_0$
			\For{$k=0$ to MaxIter}
			\State Update forward \eqref{eq:state_kinematics}:
			\[\states^{(k)}(s) = \states_0 + \int_0^{s} f(\states^{(k)}, \deformations^{(k)})\, \dif s\]
			\State Update backward \eqref{eq:transversality_condition},~\eqref{eq:costates_evolution}:
			\begin{align*}
			 &\costates^{(k)} (L_0) = - \mu_{\text{tip}} \frac{\partial \Phi_{\text{tip}} }{\partial \states} (\states^{(k)}(L_0), \states^*)  \\
			 &\costates^{(k)}(s) = \costates^{(k)}(L_0) - \int_{s}^{L_0} g(s, \states^{(k)}, \costates^{(k)},\deformations^{(k)} ) \, \dif s
			\end{align*}
			\State Update deformations \eqref{eq:update_deformations}:
			\[\deformations^{(k+1)} = \deformations^{(k)}+\eta_k\frac{\partial \hat{\controlHamiltonian}}{\partial \deformations} 
			\left(s, \states^{(k)}, \costates^{(k)}, \deformations^{(k)} \right)\Delta t\]
			\EndFor
			\State Output the final deformations as $\bar{\deformations}$
		\end{algorithmic}
		\caption{Forward-Backward Algorithm}
		\label{alg:forward_backward}
	\end{algorithm}


\section{Cosserat rod model discretization} \label{sec:numerics}

	The explicit form of the equations of motion of a planar Cosserat rod \cite{antman1995nonlinear} are as follows:
	
		\begin{align}
		\begin{split}
		\partial_{t}(\rho A \partial_t r)&=\partial_s n+ F\\
		\partial_{t}(\rho I \partial_t \theta)&=\partial_s m + \nu_1 n_2-\nu_2 n_1+ C
		\end{split}
			\label{eq:mechanics}
		\end{align}
	where $n = n_1 \a+n_2 \b$ and $m$ are internal forces and couple, respectively, and $u=(F, C)$ are external forces and couple per unit length, which are employed here as control variables.  We fix the rod base ($s=0$) at the origin while the tip ($s=L_0$) is free to move. Then, the initial \eqref{eq:initial_conditions} and boundary \eqref{eq:boundary_conditions} conditions that accompany the dynamics \eqref{eq:mechanics} are
    \newcommand{\customeqnsize}{\fontsize{9.5pt}{8pt}\selectfont}
	\begin{align}\label{eq:initial_conditions}
		\hspace{-5pt} 	\text{\customeqnsize $r(s,0) = r^{\intrinsic} (s),~\theta(s,0)=0,~ \partial_t r(s,0) =0,~\partial_t\theta(s,0)=0$}
	\end{align}
	\vspace{-21pt}
	\begin{align}\label{eq:boundary_conditions}
		\hspace{-5pt} \text{\customeqnsize $r(0,t) =0,	~\theta(0,t)=0, ~ n(L_0, t) =0,	~ m(L_0,t)=0$}
	\end{align}
	where $r^{\intrinsic}(s) = (s, 0)$ is the initial position vector.

In all our demonstrations, the arm is initially straight, undeformed and at rest. In order to mimic the tapered geometry of an actual octopus arm, we employed a rod with the variable diameter profile $\phi(s)=\phi_{\text{tip}}s+\phi_{\text{base}}(L_0-s)$. The cross section area and the second moment of the area are calculated as $A=\frac{\pi\phi^2}{4}, I=\frac{A^2}{4\pi}$. The arm dimensions of a live octopus (\textit{O. rubescens}), such as length and the diameters along the arm, were measured in a laboratory environment with the help of camera recordings. Elastic moduli of biological tissue \cite{tramacere2014structure} are used for our simulations. The simulation parameters are listed in Table \ref{tab:parameters}.

The governing equations of the Cosserat rod theory are solved numerically using our open-source, dynamic, three-dimensional (3D) simulation framework \textit{Elastica} \cite{gazzola2018forward,zhang2019modeling}. In the context of this work, we constrained all variables and motions within a prescribed plane, which acts as a fixed-point space for the dynamics. In \textit{Elastica}, the rod is decomposed into $(N+1)$ vertices hosting translational degrees of freedom ($r$), and $N$ connecting cylindrical segments hosting rotational degrees of freedom ($\theta$). All spatial operators are discretized using second-order finite-differences. The resulting discretized system of equations is evolved in time using a second-order Verlet scheme. Additional forces and torques, such as those arising from contact with objects in the environment, are included in this model as forcing terms, similar to the control $u$.  The method has been validated against a number of benchmark problems with known analytical solutions \cite{gazzola2018forward}. Moreover, it has been shown to successfully capture the dynamics of a wide range of biophysical phenomena from complex musculoskeletal architectures \cite{zhang2019modeling} and bio-hybrid robots \cite{Pagan-Diaz:2018,Aydin:2019} to artificial muscles \cite{Charles:2019} and meta-materials \cite{Weiner:2020}. Further numerical details can be found in the above references. 
	
	\begin{table}[tb]
		\centering
		\caption{Parameters}
		\hskip-10pt
		\begin{tabular}{clc}
			\hline
			\hline\noalign{\smallskip}
			Parameter & Description & Value \\
			\hline\noalign{\smallskip}
			\multicolumn{3}{c}{Numerical Simulation}\\\noalign{\smallskip}
			$L_0$ & total length of an undeformed arm [cm]& $20$ \\
			$\phi_\text{base}$ & base diameter [cm] & $2$\\
			$\phi_\text{tip}$ & tip diameter [cm] & $0.04$\\
			$E$ & Young's modulus [kPa] & $10$ \\
			$G$ & shear modulus [kPa] & $1$\\
			$\rho$ & density [kg/m$^3$] & $700$ \\
			$\gamma$ & dissipation [kg/s] & $0.01$ \\
			$N$ & discrete number of elements & $100$ \\
			$\Delta t$ & discrete time-step [s] & $10^{-5}$ \\
			\hline\noalign{\smallskip}
			\multicolumn{3}{c}{Forward-Backward Algorithm}\\\noalign{\smallskip}
			$\mu_{\text{tip}}$ & regularization parameter& $10^3$ \\
			$\eta \Delta t$ & learning rate & $0.01$\\
			$\xi$ & weight for the performance index & $10^5$\\
			\hline
		\end{tabular}
		\label{tab:parameters}
	\end{table}

\section{Numerical experiments} \label{sec:experiments}

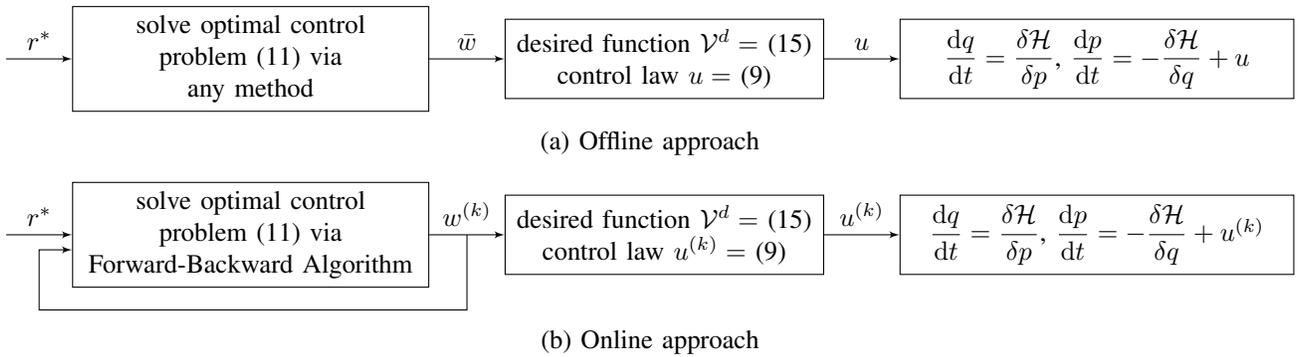
\begin{figure*}[!tb]
	\vspace{5pt}
	\begin{subfigure}{\linewidth}
		\centering
		\begin{tikzpicture}[auto, node distance=5cm,>=latex']
		\node [input, name=input] {};
		\node [block, right of=input, node distance=3.25cm] (RenderDeformation) {
			\parbox{4.5cm}{
				\centering
				solve optimal control \\
				problem \eqref{eq:optimal_control_free_end} via\\
				any method
			}
		};
		\node [block, right of=RenderDeformation, node distance=5.5cm] (ControlLaw) {
			\parbox{4cm}{
				\centering
				desired function $\potential^d=$ \eqref{eq:desired_potential}  \\
				control law $u=$ \eqref{eq:energy_shaping_control}}
		};
		\node [block, right of=ControlLaw, node distance=5.75cm] (System) {
			\parbox{5cm}{
				\centering %
				$\dfrac{\dif \states}{\dif t} = \dfrac{\delta \Hamiltonian}{\delta \momentums}$,
				$\dfrac{\dif \momentums}{\dif t} = - \dfrac{\delta \Hamiltonian}{\delta \states} + u$
			}
		};

		\draw [->] (input) -- node {$r^\target$} (RenderDeformation);
		\draw [->] (RenderDeformation) -- node[name=deformations] {$\bar\deformations$} (ControlLaw);
		\draw [->] (ControlLaw) -- node {$u$} (System);

		\end{tikzpicture}
		\caption{Offline approach}
		\label{fig:flow_chart}
	\end{subfigure}

	\vspace{0.3cm}

	\begin{subfigure}{\linewidth}
		\centering
		\begin{tikzpicture}[auto, node distance=5cm,>=latex']
		\node [input, name=input] {};
		\node [block, right of=input, node distance=3.25cm] (RenderDeformation) {
			\parbox{4.5cm}{
				\centering
				solve optimal control \\
				problem \eqref{eq:optimal_control_free_end} via\\
				Forward-Backward Algorithm\\
			}
		};
		\node [block, right of=RenderDeformation, node distance=5.5cm] (ControlLaw) {
			\parbox{4cm}{
				\centering
				desired function $\potential^d=$ \eqref{eq:desired_potential}  \\
				control law $u^{(k)}=$ \eqref{eq:energy_shaping_control}}
		};
		\node [block, right of=ControlLaw, node distance=5.75cm] (System) {
			\parbox{5cm}{
				\centering %
				$\dfrac{\dif \states}{\dif t} = \dfrac{\delta \Hamiltonian}{\delta \momentums}$,
				$\dfrac{\dif \momentums}{\dif t} = - \dfrac{\delta \Hamiltonian}{\delta \states} + u^{(k)}$
			}
		};

		\draw [->] (input) -- node[name=target]{$r^\target$} (RenderDeformation);
		\draw [->] (RenderDeformation) -- node[name=deformations] {$\deformations^{(k)}$} (ControlLaw);
		\draw [->] (ControlLaw) -- node {$u^{(k)}$} (System);

		\coordinate[below=1cm of deformations](p1);
		\coordinate[below=1cm of target](p2);
		\coordinate[below=0.2cm of target](p3);
		\coordinate[below=0.2cm of RenderDeformation.west](p4);

		\draw [->] (deformations)--(p1)--(p2)--(p3)--(p4);

		\end{tikzpicture}
		\caption{Online approach}
		\label{fig:flow_chart_online}
	\end{subfigure}

	\caption{(a) Offline approach. The block on the very left is the first step which takes the target position $r^\target$ as input and outputs target deformation $\bar\deformations$. The middle block is the second step which constructs $u$ by following the control law \eqref{eq:energy_shaping_control}. Lastly, the system will be stabilized at the target deformation once the control is applied. (b) Online approach. The two-step procedure here is similar to (a) except for the fact that the deformations are updated iteratively, and ${\deformations}^{(k)}$ is used to compute the energy shaping control $u^{(k)}$.}
	\label{fig:control_diagram}
\end{figure*}

In the following we demonstrate the capabilities of our control approach via a set of numerical experiments inspired by arm reaching motions reported in octopus' literature.

  \begin{figure*}[!ht]
  \vspace{5pt}
  \centering
  \includegraphics[width=\textwidth, trim = {0 200 150 0}, clip = true]{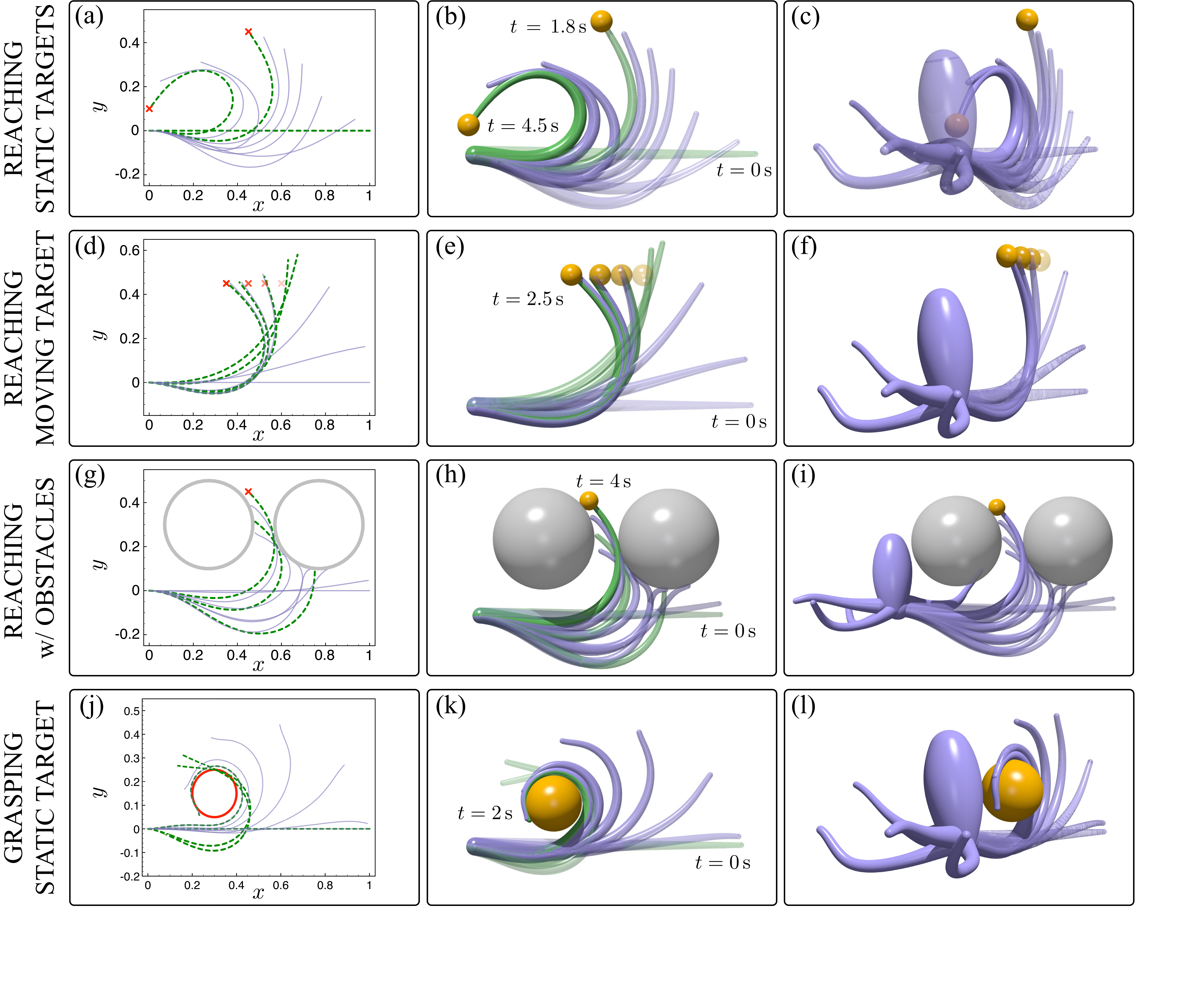}
  \caption{Arm reaching control tasks. (a-c) The arm is tasked to reach two different locations one after the other, mimicking an octopus fetching a food source and bringing it back to its mouth. (a) Targets are located at $r^\target=(9,9)$ and $(0,2)$ [cm] (axes normalized by undeformed arm length $L_0$)  and indicated as red crosses. In (b -- arm front view) and (c -- octopus 3D rendering) targets are represented as orange spheres. Optimal arm configurations are depicted in green, while actual arm shapes evolution in time are depicted in purple. (d-e) The arm is tasked to reach a moving target, initially located at $(12,9)$ [cm]. (g-i) The arm is tasked to reach a static target accounting for the presence of two identical solid spheres (grey) of $8$ [cm] diameter, and located at $(5.4,6)$ and $(15.4,6)$ [cm]. The position of the static target is $r^\target=(9,9)$ [cm]. (j-l) The arm is tasked to wrap around a static sphere of diameter $4$ [cm] centered at $(6, 3)$ [cm].} 
  \label{fig:experiments}
  \vspace{-15pt}
  \end{figure*}
	\subsection{Reaching multiple static targets}
Octopuses have been observed ~\cite{sumbre2001control, sumbre2005motor, sumbre2006octopuses} to demonstrate stereotypical reaching and fetching motion, i.e. reaching to a food source by bend propagation and bringing it back to its mouth by forming a pseudo-joint in its arm. Inspired by this, our first experiment is conceptualized for the \textit{CyberOctopus} elastic arm to mimic this kind of behavior. Given one or multiple static targets ${r}^*$, indicated as orange spheres, the goal of this numerical experiment is to reach each target with the tip of the arm one after the other. 
	The first step is using the forward-backward algorithm to calculate offline (Fig.~\ref{fig:flow_chart}) the static configuration, given each target's position. To find the static configuration that allows the tip of the arm to reach the target, we set the terminal cost in the optimal control problem \eqref{eq:optimal_control_free_end} as
	\begin{equation}
		\Phi_{\text{tip}}(\states (L_0), r^\target)=\frac{1}{2} |r^\target-r(L_0)|^2 
		\label{eq:phi_tip} 
	\end{equation}
	There is no cost associated with $\theta(L_0)$ since the angle at which the tip captures the target is not of concern. The transversality condition \eqref{eq:transversality_condition} becomes
	\begin{equation}
		\costates(L_0)=\mu_{\text{tip}}\begin{bmatrix}
		x^*-x(L_0)\\y^*-y(L_0)\\0
		\end{bmatrix} \nonumber
	\end{equation}
After computing the target configuration, we apply the explicit muscle forces and couples of \eqref{eq:control}, which smoothly bring the arm into its target shape. When the tip reaches the first target, another set of muscle forces and couples based on next target  is applied. Therefore, the arm reaches each target one by one, as shown in Fig. \ref{fig:experiments}a-c.

	\subsection{Reaching a moving target}

Next, we consider reaching a moving target so that ${r}^*$ is now an explicit function of time, mimicking the capture of a swimming prey \cite{villanueva1997swimming, villanueva2017cephalopods}. This scenario sets the stage for future investigations of capture strategies in more complex settings, for example accounting for preys' evasion maneuvers. Thus, a method that continuously updates the desired arm configuration $\bar{\states}(t)$ in response to dynamic targets becomes necessary, and we resort to the online control method of Fig.~\ref{fig:flow_chart_online}.

In this test case, the target position is displaced as $r^\target(t)=r^\target(k\Delta t)$, where $k$ is the iteration number. The target is assumed to be moving at a constant velocity of 1 [cm/s], towards the $-\mathsf{e}_1$ direction. It is to be noted that the controller for the arm does not know the velocity explicitly, instead it is assumed to know the position of the target at each time. This can be justified since the octopus can use visual cues and chemical signals to estimate the location of the prey. As can be seen in Fig.~\ref{fig:experiments}d-f, the tip of the arm catches the moving target, gradually morphing though a sequence of desired shapes.

\subsection{Reaching with obstacles}\label{sec:exp_obstacles}
Challenged with physical constraints, octopuses are known to adapt to the environment to accomplish complex tasks like reaching to a target \cite{richter2015octopus}, or solving puzzles \cite{richter2016pull}. Here, we consider the presence of solid obstacles to mimic an octopus operating in an anisotropic environment. The target is assumed to be static. We follow the method described in Sec. \ref{sec:desired_potential_energy} to find optimal static configurations that respects the inequality constraints associated with hard boundaries, here represented by two spheres located in the arm plane. Thus, the inequality constraints are
	\begin{equation*}
		\Psi_j(\states(s))=\left(\frac{\phi_j+\phi(s)}{2}\right)^2-|r_j-r(s)|^2,\quad j=1,2
	\end{equation*}
	where $\phi_j$ is the diameter of the $j$-th sphere and ${r}_j$ is its center position. The online control method (Fig. \ref{fig:flow_chart_online}) is then applied to calculate the energy-shaping control. The results of algorithm and simulations are shown in Fig.~\ref{fig:experiments}g-i, which illustrate how the tip avoids the boundaries as the arm complies with the obstacles, sliding through them to finally reach the target.

\subsection{Grasping an object}
For the final experiment, a target object is provided for the octopus arm to grasp. The running cost is designed as
	\begin{equation*}
		\Phi_{\text{grasp}}(\states(s)) = \text{dist}(\Omega, r(s))
	\end{equation*}
where $\Omega$ denotes the boundary of the object and $\text{dist}(\cdot, \cdot)$ calculates the distance between the boundary and the point $r(s)$. This object is also treated as an obstacle, modeled as an inequality constraint $\Psi$, as in Sec.~\ref{sec:exp_obstacles}.  
We choose the following weight function
	\begin{equation*}
		\mu_{\text{grasp}}(s)= \mu_{\text{tip}} \chi_{_{[0.4L_0,L_0]}} (s) 
	\end{equation*}
where $\chi_{_{[0.4L_0,L_0]}} (\cdot)$ denotes the characteristic function of $[0.4L_0,L_0]$. The weighted running cost together with inequality constraint causes the distal portion of the arm, starting from $s = 0.4L_0$, to grasp the target without penetrating it. 
The results of the energy shaping control law are illustrated in Fig.~\ref{fig:experiments}j-l.  
\color{black}

	\begin{figure}[!tb]
	\vspace{5pt}
		\centering
		\includegraphics[width=0.9\columnwidth , trim = {10 0 0 0}, clip = false]{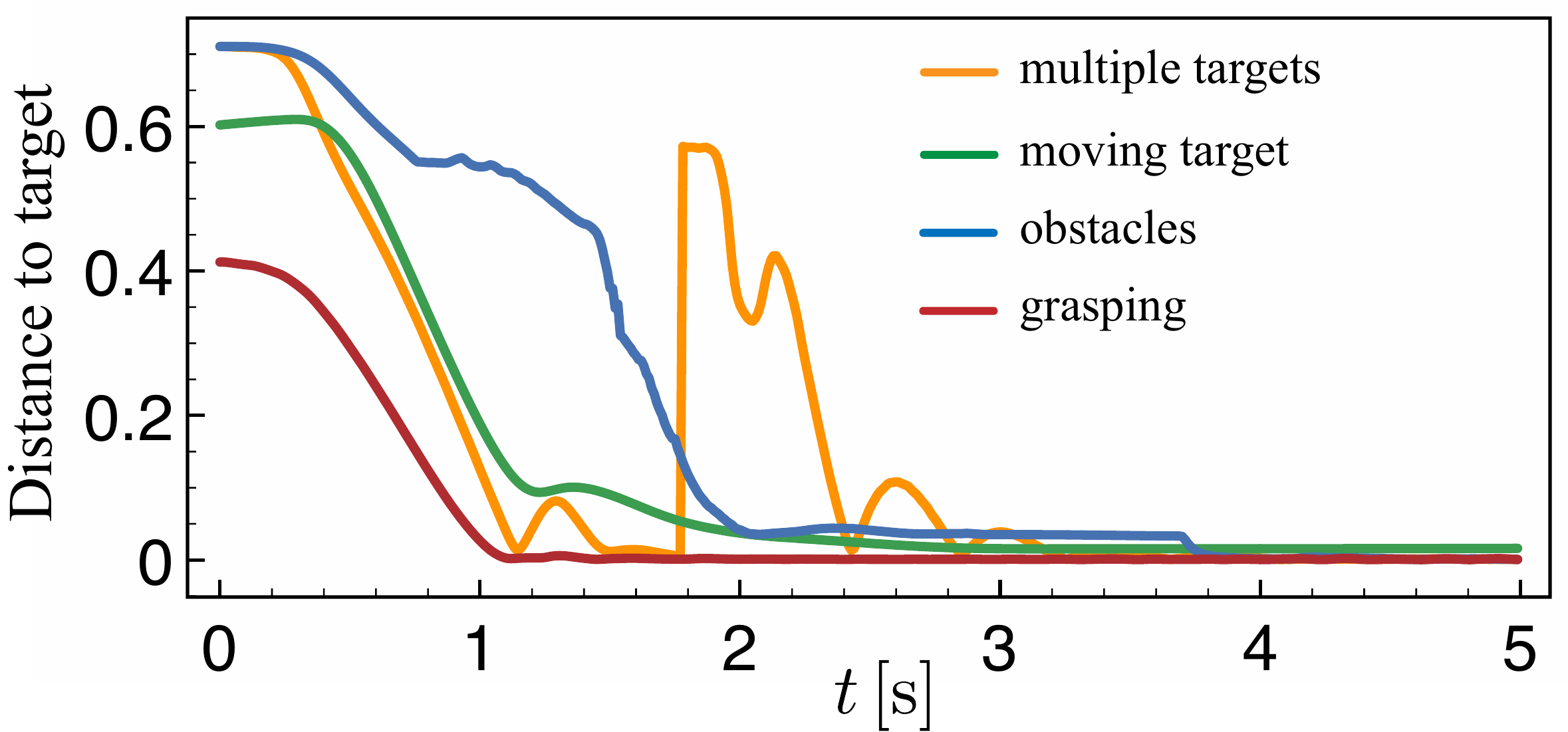}
		\caption{Time-series plot of the distance to target (normalized by $L_0$). Plotted here is the distance between the arm tip and the target positions for experiments A-C, and the averaged (weighted by $\mu_{\text{grasp}}$) distance between the arm and the boundary of the target for experiment D. For experiment A (orange line), the change of target at $1.8$ [s] is reflected by the jump. For all four cases, the distance to target smoothly approaches zero, indicating a stable equilibrium of the system.  \color{black}} 		
		\label{fig:distance}
		\vspace{-15pt}
	\end{figure}
	
\begin{remark}
In order to better understand the temporal performance of our control method, we plot the distance between the arm and the target position in Fig.~\ref{fig:distance}. Some Reinforcement Learning (RL) and adaptive control based algorithms are known to result in oscillations around the target. Compared with the manipulator results of RL based methods \cite{tassa2018deepmind, naughton2020elastica}, our proposed energy shaping control method offers the system a stable equilibrium, as well as fast computation of control. \color{black}
\end{remark}

\section{Conclusion and Future Work} \label{sec:conclusion}

	In this paper, we have used the Cosserat rod theory to model a flexible octopus arm in a plane. Hamiltonian formulation of the dynamics of the rod is exploited to synthesize an energy-shaping control law that stabilizes the rod to a predefined deformed state. We have shown that an optimal control formulation yields a systematic way to compute desired static configuration. This also enables us to tackle obstacles. An iterative forward-backward algorithm is proposed so that it can be used online to calculate the energy-shaping control in the dynamic simulation of the rod. Numerical results demonstrate efficacy of this control scheme. As a direct extension, this idea can be applied to the general 3D case. In this work, a simplistic model of actuation is assumed. Future work will consider more realistic muscle actuation models, to solve manipulation problems in a biophysically realistic fashion.

\bibliographystyle{IEEEtran}
\bibliography{reference}

\begin{thebibliography}{10}
\providecommand{\url}[1]{#1}
\csname url@rmstyle\endcsname
\providecommand{\newblock}{\relax}
\providecommand{\bibinfo}[2]{#2}
\providecommand\BIBentrySTDinterwordspacing{\spaceskip=0pt\relax}
\providecommand\BIBentryALTinterwordstretchfactor{4}
\providecommand\BIBentryALTinterwordspacing{\spaceskip=\fontdimen2\font plus
\BIBentryALTinterwordstretchfactor\fontdimen3\font minus
  \fontdimen4\font\relax}
\providecommand\BIBforeignlanguage[2]{{%
\expandafter\ifx\csname l@#1\endcsname\relax
\typeout{** WARNING: IEEEtran.bst: No hyphenation pattern has been}%
\typeout{** loaded for the language `#1'. Using the pattern for}%
\typeout{** the default language instead.}%
\else
\language=\csname l@#1\endcsname
\fi
#2}}

\bibitem{Laschi:2012}
C.~Laschi, \emph{et~al.}, ``Soft robot arm inspired by the octopus,''
  \emph{Advanced Robotics}, vol.~26, no.~7, pp. 709--727, 2012.

\bibitem{min2019softcon}
S.~Min, \emph{et~al.}, ``Softcon: Simulation and control of soft-bodied animals
  with biomimetic actuators,'' \emph{ACM Transactions on Graphics}, vol.~38,
  no.~6, 2019.

\bibitem{sumbre2001control}
G.~Sumbre, \emph{et~al.}, ``Control of octopus arm extension by a peripheral
  motor program,'' \emph{Science}, vol. 293, no. 5536, pp. 1845--1848, 2001.

\bibitem{sumbre2005motor}
------, ``Motor control of flexible octopus arms,'' \emph{Nature}, vol. 433,
  no. 7026, p. 595, 2005.

\bibitem{levy2017motor}
G.~Levy, \emph{et~al.}, ``Motor control in soft-bodied animals: the octopus,''
  in \emph{The Oxford Handbook of Invertebrate Neurobiology}, 2017.

\bibitem{antman1995nonlinear}
S.~S. Antman, \emph{Nonlinear Problems of Elasticity}.\hskip 1em plus 0.5em
  minus 0.4em\relax Springer, 1995.

\bibitem{simo1988hamiltonian}
J.~C. Simo, J.~E. Marsden, and P.~S. Krishnaprasad, ``The {H}amiltonian
  structure of nonlinear elasticity: the material and convective
  representations of solids, rods, and plates,'' \emph{Archive for Rational
  Mechanics and Analysis}, vol. 104, no.~2, pp. 125--183, 1988.

\bibitem{dichmann1996hamiltonian}
D.~Dichmann, J.~Maddocks, and R.~Pego, ``Hamiltonian dynamics of an elastica
  and the stability of solitary waves,'' \emph{Archive for Rational Mechanics
  and Analysis}, vol. 135, no.~4, pp. 357--396, 1996.

\bibitem{van2000l2}
A.~van~der Schaft, \emph{L2-gain and passivity techniques in nonlinear
  control}.\hskip 1em plus 0.5em minus 0.4em\relax Springer, 2000.

\bibitem{ortega2001putting}
R.~Ortega, \emph{et~al.}, ``Putting energy back in control,'' \emph{IEEE
  Control Systems Magazine}, vol.~21, no.~2, pp. 18--33, 2001.

\bibitem{franco2020energy}
E.~Franco and A.~Garriga-Casanovas, ``Energy-shaping control of soft continuum
  manipulators with in-plane disturbances,'' \emph{The International Journal of
  Robotics Research}, p. 0278364920907679, 2020.

\bibitem{gazzola2018forward}
M.~Gazzola, \emph{et~al.}, ``Forward and inverse problems in the mechanics of
  soft filaments,'' \emph{Royal Society Open Science}, vol.~5, no.~6, p.
  171628, 2018.

\bibitem{zhang2019modeling}
X.~Zhang, \emph{et~al.}, ``Modeling and simulation of complex dynamic
  musculoskeletal architectures,'' \emph{Nature Communications}, vol.~10,
  no.~1, pp. 1--12, 2019.

\bibitem{bretl2014quasi}
T.~Bretl and Z.~McCarthy, ``Quasi-static manipulation of a {K}irchhoff elastic
  rod based on a geometric analysis of equilibrium configurations,'' \emph{The
  International Journal of Robotics Research}, vol.~33, no.~1, pp. 48--68,
  2014.

\bibitem{till2017elastic}
J.~Till and D.~C. Rucker, ``Elastic stability of cosserat rods and parallel
  continuum robots,'' \emph{IEEE Transactions on Robotics}, vol.~33, no.~3, pp.
  718--733, 2017.

\bibitem{takegaki1981new}
M.~Takegaki and S.~Arimoto, ``A new feedback method for dynamic control of
  manipulators,'' \emph{Journal of Dynamic Systems, Measurement, and Control},
  vol. 103, no.~2, pp. 119--125, 1981.

\bibitem{mekarapiruk1997optimal}
W.~Mekarapiruk and R.~Luus, ``Optimal control of inequality state constrained
  systems,'' \emph{Industrial \& Engineering Chemistry Research}, vol.~36,
  no.~5, pp. 1686--1694, 1997.

\bibitem{morrison1962multiple}
D.~D. Morrison, J.~D. Riley, and J.~F. Zancanaro, ``Multiple shooting method
  for two-point boundary value problems,'' \emph{Communications of the ACM},
  vol.~5, no.~12, pp. 613--614, 1962.

\bibitem{healey2006straightforward}
T.~J. Healey and P.~G. Mehta, ``Straightforward computation of spatial
  equilibria of geometrically exact cosserat rods,'' \emph{International
  Journal of Bifurcation and Chaos}, vol.~15, no.~3, pp. 949--965, 2005.

\bibitem{mcasey2012convergence}
M.~McAsey, L.~Mou, and W.~Han, ``Convergence of the forward-backward sweep
  method in optimal control,'' \emph{Computational Optimization and
  Applications}, vol.~53, no.~1, pp. 207--226, 2012.

\bibitem{taghvaei2017regularization}
A.~Taghvaei, J.~W. Kim, and P.~Mehta, ``How regularization affects the critical
  points in linear networks,'' in \emph{Advances in Neural Information
  Processing Systems}, 2017, pp. 2502--2512.

\bibitem{tramacere2014structure}
F.~Tramacere, \emph{et~al.}, ``Structure and mechanical properties of octopus
  vulgaris suckers,'' \emph{Journal of The Royal Society Interface}, vol.~11,
  no.~91, p. 20130816, 2014.

\bibitem{Pagan-Diaz:2018}
G.~Pagan-Diaz, \emph{et~al.}, ``Simulation and fabrication of stronger, larger,
  and faster walking biohybrid machines,'' \emph{Advanced Functional
  Materials}, vol.~28, no.~23, p. 1801145, 2018.

\bibitem{Aydin:2019}
O.~Aydin, \emph{et~al.}, ``Neuromuscular actuation of biohybrid motile bots,''
  \emph{Proceedings of the National Academy of Sciences}, vol. 116, no.~40, pp.
  19\,841--19\,847, 2019.

\bibitem{Charles:2019}
N.~Charles, M.~Gazzola, and L.~Mahadevan, ``Topology, geometry, and mechanics
  of strongly stretched and twisted filaments: solenoids, plectonemes, and
  artificial muscle fibers,'' \emph{Physical Review Letters}, vol. 123, p.
  208003, 2019.

\bibitem{Weiner:2020}
N.~Weiner, \emph{et~al.}, ``Mechanics of randomly packed filaments - the "bird
  nest" as meta-material,'' \emph{Journal of Applied Physics}, vol. 127, no.~5,
  p. 050902, 2020.

\bibitem{sumbre2006octopuses}
G.~Sumbre, \emph{et~al.}, ``Octopuses use a human-like strategy to control
  precise point-to-point arm movements,'' \emph{Current Biology}, vol.~16,
  no.~8, pp. 767--772, 2006.

\bibitem{villanueva1997swimming}
R.~Villanueva, C.~Nozais, and S.~v. Boletzky, ``Swimming behaviour and food
  searching in planktonic octopus vulgaris cuvier from hatching to
  settlement,'' \emph{Journal of Experimental Marine Biology and Ecology}, vol.
  208, no. 1-2, pp. 169--184, 1997.

\bibitem{villanueva2017cephalopods}
R.~Villanueva, V.~Perricone, and G.~Fiorito, ``Cephalopods as predators: a
  short journey among behavioral flexibilities, adaptions, and feeding
  habits,'' \emph{Frontiers in Physiology}, vol.~8, p. 598, 2017.

\bibitem{richter2015octopus}
J.~N. Richter, B.~Hochner, and M.~J. Kuba, ``Octopus arm movements under
  constrained conditions: adaptation, modification and plasticity of motor
  primitives,'' \emph{Journal of Experimental Biology}, vol. 218, no.~7, pp.
  1069--1076, 2015.

\bibitem{richter2016pull}
------, ``Pull or push? octopuses solve a puzzle problem,'' \emph{PloS One},
  vol.~11, no.~3, 2016.

\bibitem{tassa2018deepmind}
Y.~Tassa, \emph{et~al.}, ``Deepmind control suite,'' \emph{arXiv preprint
  arXiv:1801.00690}, 2018.

\bibitem{naughton2020elastica}
N.~Naughton, \emph{et~al.}, ``Elastica: A compliant mechanics environment for
  soft robotic control,'' in \emph{Conference on Robot Learning, submitted},
  2020.

\end{thebibliography}

\appendices
\renewcommand{\thelemma}{A-\arabic{section}.\arabic{lemma}}
\renewcommand{\thetheorem}{A-\arabic{section}.\arabic{theorem}}
\renewcommand{\theequation}{A-\arabic{equation}}
\renewcommand{\thedefinition}{A-\arabic{definition}}
\setcounter{lemma}{0}
\setcounter{theorem}{0}
\setcounter{equation}{0}

\end{document}